\documentclass[10pt,letterpaper]{article}
\usepackage[top=0.85in,left=2.75in,footskip=0.75in]{geometry}

\usepackage{amsmath,amssymb}

\usepackage{changepage}

\usepackage{textcomp,marvosym}

\usepackage{cite}

\usepackage{nameref,hyperref}

\usepackage[right]{lineno}

\usepackage[nopatch=eqnum]{microtype}
\DisableLigatures[f]{encoding = *, family = * }

\usepackage[table]{xcolor}

\usepackage{array}

\newcolumntype{+}{!{\vrule width 2pt}}

\newlength\savedwidth

\raggedright
\usepackage[aboveskip=1pt,labelfont=bf,labelsep=period,justification=raggedright,singlelinecheck=off]{caption}

\makeatletter
\renewcommand{\@biblabel}[1]{\quad#1.}
\makeatother

\usepackage{lastpage,fancyhdr,graphicx}
\usepackage{epstopdf}
\fancyheadoffset[L]{2.25in}
\fancyfootoffset[L]{2.25in}
\begin{document}
\vspace*{0.2in}

\begin{flushleft}
{\Large
\textbf\newline{What Motivates People to Trust `AI' Systems?} 
}
\newline
\\
Nanna Inie\textsuperscript{1,2},
\\
\bigskip
\textbf{1} Department of Computer Science/Center for Computing Education Research (CCER), IT University of Copenhagen, Copenhagen, Denmark
\\
\textbf{2} Paul G. Allen School of Computer Science \& Engineering, University of Washington, Seattle, Washington, USA
\\
\bigskip

%
%





nans@itu.dk

\end{flushleft}
\section*{Abstract}
Companies, organizations, and governments across the world are eager to employ so-called `AI' (artificial intelligence) technology in a broad range of different products and systems. The promise of this cause célèbre is that the technologies offer increased automation, efficiency, and productivity --- meanwhile, critics sound warnings of illusions of objectivity, pollution of our information ecosystems, and reproduction of biases and discriminatory outcomes. This paper explores patterns of motivation in the general population for trusting (or distrusting) `AI' systems. Based on a survey with more than 450 respondents from more than 30 different countries (and about 3000 open text answers), this paper presents a qualitative analysis of current opinions and thoughts about `AI' technology, focusing on reasons for trusting such systems. The different reasons are synthesized into four \textit{rationales} (lines of reasoning): the \textit{Human favoritism rationale}, the \textit{Black box rationale}, the \textit{OPSEC rationale}, and the \textit{`Wicked world, tame computers' rationale}. These rationales provide insights into human motivation for trusting `AI' which could be relevant for developers and designers of such systems, as well as for scholars developing measures of trust in technological systems.



\section*{Introduction}
`AI' (artificial intelligence) --- probabilistic automation\footnote{The denomination ``artificial intelligence'' is poorly defined, and does not refer to a coherent set of technologies. In general, discussions of technologies called `AI' become more lucid and thus productive when we speak about the automation of specific tasks \cite{langer2022look}. In the case of this research, the fictitious systems presented to the participants vary in their task domain, but they  are all imagined to be built on statistical analysis of large datasets. Therefore, these systems are referred to collectively as ``probabilistic automation''.} --- is increasingly embedded into more and more processes and existing technologies. The promises of these technologies are ambitious and lofty: businesses and economies stand to flourish through increased automation (and therefore, productivity) \cite{mckinseyPromiseChallenge}, creativity and decision-making will be augmented \cite{eyStartTruly}, and education will be accessible to almost everyone \cite{promisesEducation}. Meanwhile, we are also warned about risks of reproduction of discriminatory bias in everything from search results to social and justice systems, exploitation of workers and data theft \cite{bender23testimony}, and pollution of our information ecosystems by synthetic media \cite{shah2023envisioning}. What is the average person to expect and believe about probabilistic automation systems? What \textit{do} they expect and believe?

We currently do not have much empirical knowledge about what contributes to people’s trust in probabilistic automation systems, and we are lacking ways of assessing or measuring trust (by people) and trustworthiness (of systems) \cite{hallowell2022don}. Without such knowledge, it is difficult to move `from principles to practice' \cite{duenser2023trust}, because developers and designers of probabilistic automation systems do not know which dimensions of the technologies conjure more or less trust in systems and why. 

This paper builds on the qualitative responses from a survey of more than 450 people from all over the world. The participants were shown eight pairs of two (fictitious) probabilistic automation systems and asked which one they would be more likely to trust and why. The analysis was guided by the following research question: \textbf{What motivates people to trust probabilistic automation (or `AI') systems?}

Almost 3000 open text responses were subjected to a thematic analysis, which resulted in distilled themes of \textit{incentives} and \textit{counterincentives} for people's willingness to trust probabilistic automation systems, and \textit{rationales} (i.e., lines of reasoning) that impact people's attribution of trustworthiness to the system. 
These themes and rationales are relevant to designers and developers of probabilistic automation systems who wish to know more about how such systems might be evaluated and received by the broader public. They are also relevant to scholars who work on constructing measurements of trust in probabilistic automation systems \cite{ueno2022trust}.


\section{Related work}

\subsection{Motivation, rationale, and the Technology Acceptance Model (TAM)}

Motivation can be broadly defined as the reasons that underlie behavior or that drive us to do something, e.g.,  \cite{broussard2002relationship, guay2010intrinsic, lai2011motivation}. Understanding motivation and motivations is inherently important in Human-Computer Interaction research, because motivation drives the reasoning for why we engage with and continue using some systems and not others \cite{peters2018designing}. This research is also concerned with people's \textit{rationale}; the reasons and justifications for why their decision makes sense and is in accordance with their underlying motivation. For example, someone who is \textit{motivated} to loose weight might choose a high-fat-low-carb diet based on the rationale that this approach is more likely to help them achieve their goal, while someone else might choose a low fat diet. The motivation is the same, but the logical reasoning differs based on a person's knowledge, previous experiences, and personal beliefs.

The terms motivation and rationale are closely tied together in the intention to and reason for interacting with products and systems. In the Technology Acceptance Model (or TAM), a person’s \textit{perception of the usefulness} and \textit{ease of use} of a system determines the person’s \textit{attitude toward using the system}, which in turn determines whether the person forms an \textit{intention to use the system} and actually uses it \cite{davis1989perceived, davis1993user}. Meta-studies have shown that attitude is essential in explaining the experiential component of human–computer interactions, that is, perceived enjoyment has a
strong effect on attitude, and considerably stronger than the effects of perceived usefulness and perceived ease of use on attitude \cite{hornbaek2017technology}. Two factors are particularly interesting in relation to probabilistic automation systems and laypeople: First: Most laypeople have not yet had a change of enough repeated experiences with probabilistic automation systems and large language models to form an grounded opinion of \textit{perceived usefulness} or \textit{ease of use} of such systems. Furthermore, many of the models are developed so rapidly that what is state of the art today will be outdated in 4-6 months. It is unclear what the influence of this momentum means for the progressive development of trust \cite{hoffman2018metrics}, and we simply do not know on which bases people currently form their opinions and attitudes towards probabilistic automation systems. Second: Probabilistic automation algorithms are often built into another system, adding some, but not all, functionality to a separate user interface. The probabilistic automation technologies themselves are often called ``black boxes''; the information architecture of the systems are obscure, opaque, and hidden from human comprehension \cite{von2021transparency, castelvecchi2016can, rai2020explainable}. This further complicates the matter of evaluating the systems -- both for their users and for researchers that study experiences of users. How can we make sure we ask people to evaluate the same thing if we are not sure the users understand what we ask them to evaluate? These two factors makes it especially relevant to (re-)investigate the rationales behind people's attitudes towards probabilistic automation systems, because we can not assume that these rationales are the same as for other technologies.


\subsection{Trust and probabilistic automation systems}
In a general sense, trust can be defined as \textit{``the willingness of a party to be vulnerable to the actions of another party based on the expectation that the other will perform a particular action important to the trustor, irrespective of the ability to monitor or control that other party''} \cite{mayer1995integrative}. Vulnerability is a key element in this definition --- in the interaction with a system, we make ourselves vulnerable to the consequences of an action. Trust involves depending on someone or something to do something, and, describes, at the same time, someone's attitude towards that dependence: \textit{``These two aspects are constitutively connected: one cannot identify acts of trusting independently of the attitude of trust they are done with''} \cite{faulkner2014practical}. Because of this interconnectedness, it makes sense to study someone's attitude towards a system independently of studying their actions and interactions with systems.

Hoffman et al. presented a study of metrics for explainable `AI' (`XAI'), and defined `whether the user's trust and reliance on the AI are appropriate' as one of the main themes \cite{hoffman2018metrics}. They presented results from a literature survey of trust scales, and distilled two essential questions for measuring trust in probabilistic automation systems:
\begin{itemize}
    \item Do you trust the machine's outputs? (which measures, or at least addresses, trust) and
    \item Would you follow the machine's advice? (which addresses reliance)
\end{itemize}
The trust measuring scales that Hoffman et al. identified in their survey assumed that a participant has had considerable experience using the system. As described in the previous section, this is difficult to assume in general, and even more difficult for probabilistic automation systems. Pragmatically, it is diffucult to conduct a priori UX evaluations of such systems because they can not be deployed without exposing people to their potential risks, thereby creating a catch-22 problem. Furthermore, the systems evolve constantly --- in fact, that is the nature of probabilistic automation and machine learning systems. Their algorithms are updated and their output changes based on new data. Moreover, Hoffman et al. points our that the existing scales for measuring trust all treat trust as a static concept, and have no way of accounting for trust that develops and changes over time \cite{hoffman2018metrics}. For the purpose of creating `XAI', they recommend that multiple measures are taken over time to track how users maintain trust \cite{hoffman2018metrics}. The work presented in this paper adds to the existing trust scales the standpoint that trust is not a binary measurement. We would not expect a user to either trust or not trust a system --- and neither would we expect 1000 users to understand the question in the same way. How these challenges were addressed in the design of the survey in this study is expanded upon in the Methodology section.

Bach et al. conducted a recent literature review of user trust in probabilistic automation systems and identified three main themes on the influences on user trust: socio-ethical considerations, technical and design features, and the user's own characteristics \cite{bach2022systematic}. Based on the investigation of 23 previous studies, they found that trust is highly context and user dependent, and the paper highlights the importance of understanding how trust varies between different user groups and attributes. The user characteristics emphasized in their study are inherent characteristics (such as personality traits), acquired characteristics (e.g., education and experience with AI), user attitudes (e.g., acceptance, expectations, and perceptions), and external variables (e.g., initial interactions, cognitive load, and the time spent with the system).

\subsection{\textit{Trustworthiness} and probabilistic automation systems}

Duenser et al. declare that it is important that probabilistic automation systems are recognized as sociotechnical systems, \textit{``where the people involved in designing, developing, deploying, and using the system are as important as the system for determining whether it is trustworthy''} \cite{duenser2023trust}. They highlight that `trust' and `trustworthiness' are often not clearly separated (or defined) in research, and emphasize the need for empirically investigating the relationship between the two. 

Mayer et al. \cite{mayer1995integrative} defined perceived trustworthiness as a function of three things. First, \textit{ability}, or the extent to which the trustor perceives the trustee to have the ability to successfully carry out the tasks and responsibilities expected in the relevant domain or context. Secondly, \textit{integrity}, which refers to the extent to which the trustee is perceived to adhere to a set of normative values and principles. The integrity of the trustee is evident when the trustor believes the trustee to be fair, consistent, and reliable in terms of doing what they say they will do (i.e., is dependable). Third, \textit{benevolence}, the extent to which the trustee is perceived to be caring, wanting to help, and having good intentions.

Relating to technology specifically, Baldassarre et al. define \textit{explainability}, \textit{fairness}, \textit{security}, and \textit{privacy} as key components of developing trustworthy `AI'. The goal of their work is to establish actionable guidelines to support stakeholders in `addressing TAI (trustworthy `AI') principles' \cite{baldassarre2024polaris}. The resulting framework is tested in a small-scale empirical case study with stakeholders, but the effect on \textit{end users} is unknown.

One recent study by Mehrotra et al. \cite{mehrotra2024integrity} suggests that \textit{honesty}, \textit{transparency}, and \textit{fairness about risks} are key elements to elicit trust in users due to these elements increasing the \textit{integrity} of the system. Based on a user evaluation of a calorie-predicting probabilistic automation system, they find `fairness about bias' to be the strongest influence on quantitative trust scores in that use case.


\section{Methodology}
This work explores the following research question: \textbf{What motivates people to trust probabilistic automation (or `AI') systems?}
In this research question, \textit{motivation} is understood as the `attitude towards using a system', or a person's `affective response' to a system, rather than their actual use \cite{davis1993user}.

Because the intention was to reach a broad audience from all over the world, a qualitative survey-based approach was used. The objective was to obtain enough qualitative data to be able to inductively form meaningful evidence for the individual themes. It is worth recognizing that this type of qualitative data (open text survey responses), while offering deeper knowledge than multiple choice-based based questionnaires, has limitations in the level of detail we can expect to obtain (compared to, e.g., interviews). They do, however, serve the purpose of offering insights from a very broad set of respondents.

\subsection{Survey design and product pairs}
The survey was distributed in SurveyXact during November and December 2023. It was centered around eight pairs of fictional systems based on some form of (relatively vague) probabilistic automation technology, giving 16 systems total (an overview of the systems is shown in Table \ref{tab:systems}). 

\begin{table} 
    \centering
    \renewcommand{\arraystretch}{1.1}%
    \small
    \begin{tabular}{l|m{4cm}|m{4cm}}
        \textbf{Product genre} & \textbf{Product A} & \textbf{Product B} \\ \hline
        \textbf{Recommender systems} & re-Commender: recommends restaurants based on yours and other users' history & IntelliTrade: Analyzes stocks and news and recommends investment opportunities. \\ \hline
        \textbf{Personal assistant} & MonAI Maker: Helps plan personal finances based on bank statements. & Cameron: An email organizing and responding system. \\ \hline
        \textbf{Autonomous vehicles} & HaulIT: Truch that transport goods from A to B. & Commuter: Driverless bus that drives people from A to B overnight. \\ \hline
        \textbf{Drones} & AquaSentinel AI-MAR: Drone that monitors enemy seas. & AI Scan Guards: Drone that monitors land territories.  \\ \hline
        \textbf{Legal recommendations} & Judy: A system to provide help to jury members based on previous court cases. & JurisDecide: A system to help legal professionals find relevant information. \\ \hline
        \textbf{Online health diagnoses} & MindHealth: Analyzes a person's speech for signs of dementia. & DermAI Scan: Analyzes a photo for dermatological conditions. \\ \hline
        \textbf{AI Tutor} & Lingua: Language learning service that works with both speech and text. & MentorMe: Academic chatbot to provide guidance on academic topics. \\ \hline
        \textbf{Assisted shopping} & WardrobEase: Automatically restocks a person's wardrobe based on user input and previous purchases. & Shoppr: System that creates meal plans, shopping lists, and completes grocery shopping based on dietary restrictions. \\ \hline
    \end{tabular}
    \vspace{2mm}
    \caption{Overview of the different probabilistic automation-based products and their genres. Full descriptions of the systems are provided in the appendix. }
    \label{tab:systems}
\end{table}

For each system we wrote a short ``pitch'' (less than 80 words), briefly describing the features of the product (the descriptions can be found in the appendix, Tables \ref{tab:productdescriptions1}-\ref{tab:productdescriptions4}. The goal of these pitches was to give a sense of the functionality of the product without being more technical than one would expect in a news article or popular literature description of a product. It has been found that language affects how people perceive and evaluate probabilistic automation technology \cite{hunger2023unhype, langer2022look}. To mitigate language effects, we created four different descriptions of each system: two descriptions that were more technical, and two that were closer to what one would expect to find in ``AI hyped'' news articles \cite{hunger2023unhype, MysteryHype}. Participants were randomly exposed to one of the descriptions. As we show in section \ref{sec:cautionrationale}, the descriptions actually had an impact on some respondents' evaluation. 

Joint evaluation (asking participants to pick one of two systems) was used instead of single evaluation (asking participants to evaluate one system at a time, because cognitive research has shown that joint evaluation can make it easier for people to evaluate ``difficult-to-evaluate attributes'' \cite{hsee1999preference}. Joint evaluation can help a person choose an objectively more valuable option, while the utility of different attributes might be underestimated in single evaluation \cite{hsee2004distinction}.

The products were paired in genres of similar utility (for instance, ``recommender systems'' or ``online health diagnostics''), to enable apples-to-apples comparisons. The selection of systems was made to cover a broad spectrum of applications of probabilistic automation systems, from the more mundane (such as restaurant recommendations) to the more extreme (such as autonomous vehicles and military drones). The goal was to elicit different types of considerations and reasoning from each survey participant, uncovering various rationales present in different decision making processes. The participant would always be asked to choose between product A and product B in one of the genres, and never between, e.g., an autonomous vehicle and a tutoring app. For each product pair, the participants were asked the following questions:

\begin{enumerate}
    \item \textbf{Thinking of yourself as a user, which of these systems are you more likely to trust?} We ask you to think about how likely you would be to trust using this system for your own purposes, assuming you would like to use the service it would provide.
    \item Tell us more about why you chose that answer (optional).
    \item \textbf{Which of these systems would you trust to give better output for its users?} Where ``better output'' means, for instance, more correct or more helpful output\footnote{The survey was also introduced with the following text: ``On the following pages, you will be introduced to 8 pairs of short pitches of technical systems. On the following pages, you will be introduced to a series of technical systems. We ask you to evaluate these systems along the following two dimensions: \textbf{How likely you would be to trust the system as a user.} We ask you to think about how likely you would be to trust using this system for your own purposes, assuming you would like to use the service it would provide. \textbf{Which systems you would trust to give better output}, where “trust“ means you think it would provide more correct or more helpful output for its users in general (if you yourself would not like to use such a product). So, even if the system didn’t provide a service relevant to you, would it be a good system for its users? Under each question you can provide further information about your rationale behind your choice – if you wish to do so.''}.
    \item Tell us more about why you chose that answer (optional).
\end{enumerate}

\noindent In this paper, we are not concerned with the choice of product, the participant made (question 1 and 3). We are analyzing only the open text answers to question 2 and 4. The questions (and introduction to the survey, cf. footnote 2) were designed to reflect the two essential questions for measuring trust identified by Hoffman et al. \cite{hoffman2018metrics} with two modifications: (1) We could not ask the user to evaluate the system's output, given that the system does not exist in reality. We therefore created a distinction between \textit{personal trust} (question 1: Would you yourself trust the system) and \textit{general trust} (question 3: Do you think the machine's output would be reliable?). (2) To make it more likely that participants would understand trust in a somewhat similar way, an introductory text as well as a short definition of trust was provided with each product pair (cf. ``where “trust“ means you think it would provide more correct or more helpful output...'').

\subsection{Participants}
Participants were recruited via the data collection platform Prolific, and compensated between £9-£15\footnote{As suggested by Prolific's standards for ``Good hourly rate''} (depending on their average time consumption) per hour for their participation. This database allowed us to create pre-screening criteria such as gender, age, country of residence, self-assessed socio-economic status, and ethnicity, to reach as diverse a group as possible (see Fig. \ref{age}, \ref{gender}, and \ref{continent} in the supporting information for participant demographics). All participants signed a consent form that their (anonymous) answers could be used for research purposes.

\subsection{Data analysis}
A \textbf{thematic analysis} of the responses was conducted in the software Condens. 
The first round of coding was conducted with an open-ended, inductive treatment, and covered about 500 responses (see Fig. \ref{codingexample} for a screenshot of the initial coding). The thematic analysis was focused on \textit{``identifying and interpreting key, but not necessarily all, features of the data, guided by the research question''} \cite{clarke2015thematic}. In practice, each response was read with the research question in mind: which motivation does the respondent provide for being willing to trust or not willing to trust the system? Which logic is present in the response? The names of the codes were `answers' to the research question, such as `accuracy', `reliability', or `risk of bias'.

All survey responses were read several times while initial codes were generated. The goal of the thematic analysis was to identify patterns that reflect the data for this context \cite{nowell2017thematic}, meaning the goal was to create themes and codes covering all the different responses, even if they were only represented in very few responses, or if some responses were tagged with more than one code, e.g. (codes in brackets): 
\begin{quote}
    \textit{``though I do not like the data collecting data aspect of it due to privacy concerns, I do think it will be more accurate as it will feed off data that I give it''} [privacy/surveillance] [higher accuracy] (P95, South Africa, F, 21-25)
\end{quote}
The result of the initial coding was a list of about 80 different codes at very different levels of abstraction (similarly to the responses, which were also at different level of detail and abstraction). 
After the first round of coding, the categories were beginning to saturate, meaning not many new codes were added when going through new responses. Before the second round of coding, the related work was revisited and the codes were grouped into thematic clusters, which became the basis of the distillation into five rationales. The five rationales were given a working title and working description based on the initial analysis.

After this, the remaining $\sim$2500 responses were coded more deductively, with the purpose of `testing', confirming, and elaborating the rationales. This process of coding and refining codes and themes lasted about three months of continuous analysis. If a response was not an obvious fit into either of the rationales, the overarching themes were revisited and modified, and the description of them was thus updated and elaborated throughout the coding of all 3099 responses. As a result of this process, one of the initial rationales was merged into the three others, as it became clear that there was too much overlap between categories to provide fruitful distinctions.

\subsection{``How Many Bloody Examples Do You Want?''}
Crabtree et al. wrote in their article with the above title on qualitative data and generalization: \textit{``Human activities contain their own means of generalization that cannot be reduced to extraneous criteria (numbers of observations, duration of fieldwork, sample size, etc.)''}~\cite{crabtree2013many}. Because the goal of this paper is \textit{discovery}, not measurement, the results section mostly abstains from indicating numbers of participants. First, it is not possible to meaningfully quantify how many respondents truly ``think'' or ``believe'' something -- let alone how strongly they think or believe it. This kind of information does not exist in an inherently metric space, and the results in the paper are a presentation of an abstracted analysis, not a quantifiable account of facts. Second, respondents to the survey likely base their reasoning on many things that they do not necessarily write down in their survey response. It is unfair (and, likely, incorrect) to assume that a survey response would elicit all participants' full reasoning --- and such a quantification also assumes that all participants have the same degree of literacy and similar ability to communicate in writing. Third, when the purpose is discovery and mapping, outliers are as interesting as the mean:  Even if only a single respondent reports something, it tells us something important. Quantitative approaches are therefore of reduced impact here.

For clarity purposes, not all quotes are presented verbatim and may include minor grammatical corrections in the Results section. Similarly, the specific product pair that the response comes from is usually not reported directly, as it is often either discernible from the response itself or not relevant to the analysis of rationales.


\section{Results}


The goal of the thematic analysis was to uncover both overall \textit{motivations}, that is, incentives and counterincentives for the intention to use one system over another \cite{davis1989perceived}, and \textit{rationales}, i.e., the line of reasoning that lead someone to believe that their choice is the best one given their motivation --- or, reasons that one system is more or less \textit{trustworthy} than the other. In practice, motivations and rationales are inherently closely tied together, and most of the time not easily distinguishable from one another based on a survey response. The criteria used in the analysis were as follows:
\begin{itemize}
    \item \textbf{Incentives or counterincentives:} The responses coded as incentives or counterincentives express a preference for one system over another, usually based on the assumption that the system would deliver what the description promises that it would. Responses coded as incentives or counterincentives do not provide any insight into \textit{why} the participant assumes the system to be more or less useful or accurate. Examples are (codes in brackets): 
    \begin{quote}
    \textit{``it just makes sense as an adult to have this kind of app''} [incentives: sensible] \\ (P277, M, South Africa, 46-50)
    \end{quote}
    \begin{quote}
    \textit{``I love food and being able to have different suggestions for different restaurants just makes my day.''} [incentives: personal relevance] \\ (P348, F, Japan, 36-40)
    \end{quote}
    \begin{quote}
    \textit{``I wouldn't actually trust either but the one I chose is the less offensive''} [counterincentives: aversion for the other choice] \\ (P304, F, England, 66+) .
    \end{quote}
    The responses coded as incentives or counterincentives tend to focus on the utility of the system rather than the functionality, i.e., would the respondent like to use the service the system would provide? In the choice between two systems, the participant will then choose the one they perceive as most useful or least bad by their own value system.

    \item \textbf{Rationales:} The responses tagged with different rationales provide a bit more detail into why the respondent deems the system more or less trustworthy. They offer terminology to help us understand the reasoning behind the attribution of trustworthiness, for instance: 1) On which \textit{premise} is the reasoning based, that is, what is the underlying argument? 2) What are the \textit{consequences} of the reasoning? If someone employs this reasoning, which decision would they make about a probabilistic automation system? 3) Which are the \textit{evaluation factors} of the reasoning? Which things are taken into account when evaluating the trustworthiness of the system?
\end{itemize}
Consider the following two responses:
\begin{quote}
    \textit{``Stock trading is a probability based decision. There are too many circumstances that influence stocks other than the news. I trust human based recommendations better.''} (P128, M, South Africa, 26-30)
\end{quote}
\begin{quote}
    \textit{``for JurisDecide, the people interpreting the data would be a lawyer, so they would be able to understand and fact check where - as as a Jury member who isn't at all clued up on relevant law you wouldn't even know how to question the response of the AI''} (P191, M, England, 21-25)
\end{quote}
In the first response, the participant is evaluating two recommender systems: one that recommends restaurants, and one that recommends investment opportunities. The participant is saying that they would trust the restaurant recommender system more because its input is `human-based'. We could say this is based on an underlying \textit{premise} that humans are more trustworthy than machines. The \textit{consequence} of such a premise might be that a person would prefer systems with greater involvement of humans and the \textit{evaluation factors} might be `to which degree is a human involved in the creation or fact checking of the output'?

In the second response, the participant is evaluating two systems used in legal contexts: one that is supposed to assist a jury, and one that is supposed to assist lawyers. The participant states that because the first system is going to be used by `laymen' (not legal professionals) who are not qualified to question the output, the participant would be less likely to trust the system. We could say this is based on the \textit{premise} that human \textit{experts} are more trustworthy than a machine, and the \textit{consequence} is that probabilistic automation systems are more trustworthy in circumstances where they are used by human domain experts. The \textit{evaluation factors} would therefore be the expertise of the specific users.
\\


The distinction between motivation and rationale may need further unpacking if one is interested in the psychological underpinnings of drivers of human behavior, but for the purpose of this exploratory paper, these working definitions suffice. The focus in this paper is only on motivations and rationales, people volunteered for making a choice between two products. The participants' choice of product is not reported in this paper (see [ANONYMIZED, under review] for these results).

\subsection{Motivations: incentives and counterincentives}

\subsubsection{Incentives}

The most prevalent incentives brought up for trusting one system over another were, perhaps unsurprisingly, \textbf{personal relevance} and \textbf{perceived usefulness} --- which would be anticipated from a technology acceptance model perspective \cite{davis1989perceived, davis1993user}. Although closely related, these codes varied slightly in their application; \textbf{personal relevance} was used when the participant mentioned something uniquely well suited for their personal circumstances or meeting a need they currently experience, such as: 
\begin{quote}
    \textit{``I chose the the first option because I am an investor and a trader, so the app would really help me with my trading''} (P285, F, Hungary, 41-45)
\end{quote}
\textbf{Perceived usefulness} described instances where the response was more vague or described needs of others, e.g., 
\begin{quote}
    \textit{``This would make the work of lawyers easier''} (P113, M, Netherlands, 36-40)
\end{quote}
Sometimes the estimation of usefulness was tied to a \textbf{specific product property} described in the text, such as providing real time analysis or that the system asked questions rather than providing answers.

Another prevalent code was of responses which assumed that the output of the chosen system would have \textbf{higher accuracy}, i.e., be more likely to be correct for the occasion --- which makes sense as an evaluation criteria, since the participants were asked directly which system they believed would provide `more correct' or `more helpful' output. This would have to be based on assumptions of both the data quality and algorithmic performance, since no specifics were given in the system pitches about either, and the respondents had had no opportunity to evaluate the fictitious systems' output. It is worth noting that many respondents based their reasoning on a misunderstood or outright wrong assumption of either, e.g.:
\begin{quote}
    \textit{``Likely to produce a more accurate output due to the fact that it uses existing information or data''} (P95, F, South Africa, 21-25)
\end{quote}
\begin{quote}
    \textit{``The drone looking at seas makes more sense. It's easy to build from a computer vision perspective, and hence has a higher chance of being accurate.''} (P120, M, UK, 31-35)
\end{quote}
In the first example, the participant reasons that the higher accuracy is produced by ``existing'' information or data, which one would hope would be the case for any probabilistic automation system which is not based on pure guesswork. The second example asserts that the one system is easier to build and therefore more likely to be accurate, while it is not actually explained in the system descriptions how either of these systems would be engineered.

Many respondents explained their choice by their assumption that the system would have a \textbf{larger target market}, benefiting more people (\textbf{utilitarianism}), or the system solving a more important issue (\textbf{problem importance}):
\begin{quote}
    \textit{``This app relates with the daily lives of a huge number of people. Eating out is something most people do daily''} (P2-248, M, England, 31-35)
\end{quote}
\begin{quote}
    \textit{``A good idea to help you with your mental health, an important issue today, we all need to feel in control more of this subject''} (P2-165, M, England, 61-65)
\end{quote}

A more specific instance of perceived usefulness was the incentive that the system could \textbf{augment human ability} by automating drudgery or assist informed decision making:
\begin{quote}
    \textit{``Cameron can streamline your emails so you can have time to focus on the task instead of the admin of the task.''} (P141, M, South Africa, 26-30)
\end{quote}
\begin{quote}
    \textit{``I don't really have the time and knowledge for such, so it would be a game changer for me.''} (P175, F, South Africa, 26-30)
\end{quote}
Two other prevalent incentives were the possibility of probabilistic automation systems to \textbf{expand knowledge/provide guidance} and \textbf{expand [one's] horizon}:
\begin{quote}
    \textit{``I like the assistance it gives, like the idea of AI coming up with additional ways of problem solving, I think the technology is really useful and basically an instant way of getting knowledge on various subjects and solving questions [...] also its like a personal friend helping you out''} (P24, M, UK, 66+)
\end{quote}
Many respondents also mentioned \textbf{previous (positive) experience} with similar products to be a determining factor, which makes sense based on theory of technology adaption \cite{davis1989perceived}.

Finally, a rare, but interesting, incentive was the gathering of broad data to support expert knowledge:
\begin{quote}
    \textit{``will definitely gather quality evidence to support proper medical professionals''} (P30, M, UK, 61-65).
\end{quote}
While most respondents described giving up their personal data as a negative factor, it is worth mentioning that the opposite view was also present, albeit not common.

\subsubsection{Counterincentives}

\noindent The code \textbf{aversion for the other choice} was the most prevalent of counterincentives for trusting a system, meaning the participant describes choosing a system only because the other choice was worse. This adheres to a lesser of two evils-principle, which was somewhat enforced by the joint evaluation method, because participants were forced to make a choice between two options, even if they would be unlikely to trust either of them:
\begin{quote}
    \textit{``The stock market is a fictional concept created by greedy, capitalist draco reptilian species. It's not real because money isn't a real thing.''} (P356, F, South Africa, 26-30)
\end{quote}
\begin{quote}
    \textit{``I rather have them watch the sea than land''} (P227, F, The Netherlands, 41-45)
\end{quote}
The code itself was not particularly useful in telling us about people's motivation. It was sometimes used when the response described the other choice as not being personally relevant to someone's situation: 
\begin{quote}
    \textit{``The WardrobEase app is pretty much useless for me.''} (P33, M, Israel, 51-55),
\end{quote} 
and sometimes used to describe more specific distaste of the opposite choice: 
\begin{quote}
    \textit{``The first one seems unreliable.''} (P59, F, UK, 26-30),
\end{quote} 
In the first example, the reasoning is simply the opposite of the code `personal relevance', and the code \textbf{personal irrelevance} was therefore created for this type of counterincentive. A subcategory of that is one person that mentioned that one of the systems was not relevant to them because the \textbf{technology was unavailable} to them (the system would run on a smartphone, and they did not own a smartphone), and they would therefore not be able to use the functions that the system would provide.

The code \textbf{risky, unspecified} was used when participants used the word risk without specifying how that risk might unfold or what the impact of the risk could be: 
\begin{quote}
    \textit{``More day to day and less risky''} (P228, M, UK, 51-55).
\end{quote}
\begin{quote}
    \textit{``It is more risky to rely on AI for stock trading.''} (P265, F, Australia, 51-55)
\end{quote}

Similarly to the incentive of positive experience with similar systems, \textbf{previous (negative) experience} with similar systems --- usually a chatbot --- was also mentioned as a counterincentive:
\begin{quote}
    \textit{``I wouldn't trust either of these based on previous experience with chatbot technologies which frequently invent information.''} (P298, F, UK, 26-30)
\end{quote}

And finally, one participant mentioned that the system's \textbf{output was unwanted} because they would not want the knowledge that it would provide:
\begin{quote}
    \textit{``I'm actually not mad keen on tests to see potential for serious illness in the future. Seems really scary.''} (P190, M, UK, 61-65).
\end{quote}

\vspace{1mm}

\noindent An overview of the codes under incentives and counterincentives is presented in table \ref{tab:incentives}. The next section hones in on the different rationales discovered in the analysis.

\begin{table}[]
\small 
    \centering
    \begin{tabular}{p{9cm}|p{3cm}}
         \textbf{Incentives} & \textbf{Counterincentives} \\ \hline
         personal relevance, perceived usefulness, expand knowledge/provide guidance, higher accuracy, utilitarianism, specific product property, curiosity/interest/fun/excitement, larger target market, augmented humanabilities, individualized/adaptive, efficiency, monetary value, expand my horizon, previous (positive) experiences, problem importance, novelty, user friendliness, ease of use, time saving, generalized use, sensible, ubiquitous, would gather knowledge
         
         & aversion for the other choice, risky, unspecified, previous (negative) experience, personal irrelevance, unavailable technology, NOT ease of use, output unwanted \\ 
        
    \end{tabular}
    \caption{An overview of the codes identified for \textit{incentives} (motivation to trust a probabilistic automation system) and \textit{counterincentives} (lack of willingness to trust a probabilistic automation system). The incentives and counterincentives describe overall preferences for one system over another or `behavioral intention to use a system' \cite{davis1989perceived}.} 
    \label{tab:incentives}
\end{table}

\subsection{Rationales}
The rationales describe lines of logical reasoning that appear to affect people's attribution of \textit{trustworthiness} to a probabilistic automation system. The rationales are not necessarily clearly distinct from incentives and counterincentives, but the rationales provide more details of why an individual might arrive at a certain conclusion of incentive or counterincentive.

The rationales are not mutually exclusive. There are many overlaps, and some of the responses could be categorized as part of more than one of the rationales --- or, a rationale could be more or less present in someone's line of reasoning. As an example, a few respondents directly self-prioritized their motivation in order of importance (codes in brackets):
\begin{quote}
    \textit{``Lingua: \\ 1. Low risk} [risky, unspecified] \\\textit{ 2. I love to travel.} [personal relevance] \\ \textit{3. The app would be helpful for learning the language of the area in real-time and be able to use it.''} [usefulness] \\ (P120, F, South Africa, 36-40)
\end{quote} 
The rationales are \textit{abstractions}, similar to models, meaning that they highlight some factors and obfuscate others. They compress the analysis of many responses into clear lines of logic and omit many of the real-world considerations that would come into play in actual decision making or evaluations. 

The criteria for writing the rationales were as follows:

\begin{enumerate}
    \item The rationales should distill and explain the logical basis for a belief about probabilistic automation that could lead to either more or less trust in any given system.
    \item They should be concrete enough to inform designers and developers of such systems, and abstract enough to encompass multiple system examples or scenarios
    \item They should represent a line of reasoning present in the surveyed population in a truthful way.
\end{enumerate}

\noindent Each rationale is given a name, a description of its \textit{Premise}, and a \textit{Consequences of the rationale}-section to explain what the rationale might mean for the evaluation of a probabilistic automation system in practice. Finally, the \textit{Evaluation factors} relevant to the rationale are presented to illustrate what is taken into consideration when evaluating a system. Often, the evaluation factors correspond to either criteria defined in previous work about probabilistic automation and trustworthiness, or directly to codes used in the thematic analysis.


\subsubsection{\textbf{The human favoritism rationale}: A system can not perform as well as a human} 
This rationale, its name included, is directly based on the article titled \textit{Human favoritism, not AI aversion} by Zhang et al. \cite{zhang2023human}, who found that the knowledge of the involvement of a human expert in the creation process increases a product's perceived quality, and the knowledge that ``AI'' was involved in the creation process, does not. The rationale presumes that a human in the loop increases the trustworthiness of a probabilistic automation system, and that humans should have control over the system:
\begin{quote}
    \textit{``Judy is a better app [because] it works mainly with direct instructions.''} (P263, M, UK, 26-30)
\end{quote}
The central aspect of the human favoritism rationale is the elevation of general human capacity in favor of computational capacity, rather than a distinct dislike of technology, even though in practice, the two were often observed together:
\begin{quote}
    \textit{``I'd prefer to not use AI in neither. Justice should be at the hands of people, not AI.''} (P35, Non-binary, Portugal, 21-25)
\end{quote}
The rationale was especially present in the choice between the \textit{autonomous vehicles} pair, where many people clearly favored the presence of a human. Interestingly, many respondents even described their preference for the autonomous bus (Commuter) over the autonomous truck (HaulIT) under the assumption that the bus would still have a driver, even though no driver was ever mentioned in the system description:
\begin{quote}
    \textit{``Because there will be both human and AI involved in the decision making process.''} (P151, F, South Africa, 26-30)
\end{quote}
A curious instance of the human favoritism rationale was that some respondents argued that the system itself was more trustworthy because it \textit{felt more human}:
\begin{quote}
    \textit{``The Lingua seems more human and less "bot to do specific things", i think it matters how you interact with users.''} (P157, M, Poland, 18-20).
\end{quote}
\begin{quote}
    \textit{``This app feels more like a human rather than AI outputting answers only.''} (P104, M, South Africa, 26-30)
\end{quote}

\paragraph{\textbf{Consequences of the human favoritism rationale}}
Under the human favoritism rationale, a system is evaluated as more trustworthy if a(ny) human is in charge of the creation or evaluation of output -- and, consequently, as less trustworthy if no human is present in the creation of data (for instance, if a system automatically processes data from different sources on the internet, rather than data input by a user themselves) or in the evaluation of data. The rationale causes a preference for systems which provide \textit{`passive recommendations'} rather than acting on behalf of a human: 
\begin{quote}
    \textit{``It is suggestive and therefore a bit more reliable''} (P352, F, South Africa, 26-30)
\end{quote}
An inherent risk in this rationale is the overestimation of human capability or a tendency to overlook the shortcomings of human ability. At the extreme end of this rationale lies the assumption that human performance is ideal, and that technology does not have anything to offer:
\begin{quote}
    \textit{``I am confident with language and not overwhelmed by emails so Cameron would be of no value''} (P292, F, UK, 66+)
\end{quote} 
\begin{quote}
    \textit{``I know how volatile the markets are and conduct my own research.''} (P160, F, South Africa, 31-36)
\end{quote}

\paragraph{\textbf{Evaluation factors in the human favoritism rationale}}
Factors that are relevant in a person's evaluation of a probabilistic automation system within this rationale are the person's \textit{confidence in own and others’ ability}, the degree to which the \textit{human has control over the system}, and the \textit{passiveness of the system.}


\subsubsection{\textbf{The black box rationale}: We should not trust what we can not question and explain}
The black box rationale is related to the human favoritism rationale, but slightly more particular: it stipulates that human expertise and knowledge (not just the involvement of any human) increases trustworthiness, and that probabilistic automation systems should be used by people that are qualified to evaluate their outputs. The black box rationale is of course named after the general issue of probabilistic automation operating in unpredictable ways based on unknowable calculations, e.g., \cite{castelvecchi2016can, rai2020explainable, von2021transparency}, and within this rationale, the verification of output by human experts increases trustworthiness: 
\begin{quote}
    \textit{``Justice decisions must be individualized. Help should be used by professionals only.''} (P317, M, France, 56-50)
\end{quote}
\begin{quote}
    \textit{``Dementia is scary but so are both of these apps, they just lead you up the Google diagnose path''} (P225. F, England, 66+)
\end{quote}
In many responses, `expertise' was exemplified as \textit{domain knowledge} rather than technical knowledge (whether it was the respondent themselves or imagined others):
\begin{quote}
    \textit{``I have more trust in an app that recommends restaurants than an AI app that recommends financial investments only because of my lack of literacy in finance''} (P262, M, Ontario, Canada, 51-55)
\end{quote}
The rationale was also expressed as concerns for the general public if probabilistic automation systems are deployed widely to people who overrely on their output:
\begin{quote}
    \textit{``I found [the selection between MindHealth and DermAI Scan] really difficult to choose between because I'd be concerned about their effect on the 'worried well'''} (P301, F, UK, 56-60)
\end{quote}

\paragraph{\textbf{Consequences of the black box rationale}}
A system is more trustworthy if deployed in contexts where human (domain) experts can evaluate their output. The black box metaphor for probabilistic automation algorithms is inherently tied to issues of (lack of) \textit{transparency} \cite{rai2020explainable, von2021transparency}, and trustworthiness is increased by making systems, their data, and their algorithms more understandable and explainable:
\begin{quote}
    \textit{``[I chose this system because it] contains more information of how it works''} (P314, F, UK, 51-55)
\end{quote}
\begin{quote}
    \textit{``The description of this app is more open on the data being used which gives trust''} (P373, M, The Netherlands, 46-50)
\end{quote}
A challenge inherent in this rationale is the potential for disagreements about who is an expert, or what it means to have expertise. An interesting implication of this rationale is that it would be difficult to use probabilistic automation in any educational context where the users are learners of a subject:
\begin{quote}
    \textit{``I found this choice more difficult as it was like comparing Duolingo with ChatGPT. The reason I chose the language one was because the chatbot is likely to be able to output well structured sentences and sound correct but may get the actual content wrong (i.e. the paragraphs will all read correctly but the information they contain will be wrong). If you’re not already an expert on this subject it would be hard to not believe the incorrect information.''} (P403, F, England, 26-30).
\end{quote}

\paragraph{\textbf{Evaluation factors in the black box rationale}}
With the black box rationale \textit{explainability} \cite{baldassarre2024polaris}, \textit{transparency} \cite{von2021transparency, rai2020explainable} of the system and its data, and some evaluation of the \textit{expertise of end users} are factors relevant to the evaluation of a system's trustworthiness.


\subsubsection{\textbf{The OPSEC rationale}: We should automatically be cautious about any use of `AI'} \label{sec:cautionrationale}
This rationale has been named OPSEC after the security term Operations Security, which pertains to identification of critical information, analysis of threats, analysis of vulnerabilities, assessment of risks, and application of appropriate countermeasures \cite{force2017security}. In the OPSEC rationale lies a constant alertness towards the safety of probabilistic automation systems from a standpoint that such technology inherently carries risk, and that the technology is inappropriate (untrustworthy) to use in many contexts:
\begin{quote}
    \textit{``Neither [of the systems]. Legal professionals should not use AI.''} (P393, F, UK, 31-35)
\end{quote}
\begin{quote}
    \textit{``I wouldn't trust anything driverless period.''} (P182, F, South Africa, 31-35)
\end{quote}
Many responses expressed a clear cautionary stance towards ``anything AI''. This was often focused on general skepticism towards privacy, surveillance, and (mis)use of data:
\begin{quote}
    \textit{``The first is pure totalitarian shit: easily converted into Big Brother.''} (P278, No answer, No answer, No answer)
\end{quote}
\begin{quote}
    \textit{``I wouldn't use any of them due to privacy concerns.''} (P33, M, Poland, 21-25)
\end{quote}
\begin{quote}
    \textit{``The other one is too intrusive and would need too much sensitive information to make any useful recommendations.''} (P364, M, England, 51-55)
\end{quote}
\begin{quote}
    \textit{``You don't know who is controlling MonAIMaker so you can't trust who has your personal data.''} (P310, F, Netherlands, 46-50)
\end{quote} 
--- sometimes based on intangible feelings of what constitutes `personal data':
\begin{quote}
    \textit{``Food feels somehow more personal to me than my own economic well being. I don't get it either.''} (P211, M, CA, USA, 36-40)
\end{quote}
Another prominent factor in the rationale is an orientation towards \textit{risk} --- risk was also mentioned as a prevailing \textit{counterincentive} for using probabilistic automation systems in an unspecified sense, but in the OPSEC rationale risks were specified as for instance \textit{risk of bias}, \textit{data leaks}, and \textit{concerns of dark patterns}, which in this context entails the purposeful design of the system to perform against the interest of the user, e.g.:
\begin{quote}
    \textit{``I've heard that roboadvisors sometimes experiment a bit with customers - it's NOT that they perform badly, but that they are designed to do that.''} (P228, F, Spain, 31-35)
\end{quote} 
In some cases, the rationale was tied directly to the \textit{description} of the system or its semantics: the use of words like ``AI'' or ``neural networks'' automatically decreases trust. This confirms previous quantitative evaluations such as \cite{langer2022look}, and highlights the importance of obtaining qualitative data to understand the nuances of evaluations of technology. The counterpreference for ``hype'' terminology was apparent from some of the responses to the system descriptions which were deliberately stripped of such terminology --- some respondents would simply misunderstand this to mean that these systems were more trustworthy because they did \textit{not} use `AI' or similar technology (even though the technical description was actually synonymous with `AI' technology):
\begin{quote}
    \textit{``I'm not sure I would entirely trust Cameron not to miss any important/urgent emails. However when it came to my data I'd trust it more than any AI''} (P53, M, UK, 51-55) 
\end{quote}
\begin{quote}
    \textit{``I assume from the name and description MonAIMaker uses AI therefore I wouldn't trust it at all.''} (P53, M, UK, 51-55)
\end{quote}
\begin{quote}
    \textit{``This one doesn't use neural networks\footnote{The substituted term was `weighted networks'} so it's most likely to be more accurate''} (P172, NB, South Africa, 26-30)
\end{quote}
The ``AI hype'' in popular media has caused a motif of skepticism especially connected to the way we speak about `AI' systems. One way of resisting the hype has been characterized as describing `AI' systems in terms of what they help us do, rather than what capabilities we claim they have (such as `intelligence', `thinking', or `understanding') \cite{Bender2024resisting}. Interestingly, this could lead to a covert risk that people might involuntarily `lower their parades' if a description of a system avoids hype terms. People can therefore be mislead to trust systems that actually use the exact same technology they sought to avoid in the first place.

\paragraph{\textbf{Consequences of the OPSEC rationale}}
The immediate consequence of this rationale is to automatically be cautious about any use of ‘AI’. In the OPSEC rationale, not many systems can be deemed trustworthy if they use probabilistic automation technologies. One common reason given for the choice of one system over another was that the system had ``lower stakes'' --- that the \textit{impact in case of failure} of the system would be lower. Lower stakes would often  mean less risk of damage to humans:
\begin{quote}
    \textit{``HaulIT doesn't transport humans, who can get injured or killed if Commuter makes a mistake.''} (P140, F, South Africa, 41-45)
\end{quote}
But lower stakes also translated to, for instance, monetary risk or risk of damage to interpersonal relationships:
\begin{quote}
    \textit{``Lower stakes - only deals with hobbies/past times as opposed to finances''} (P13, M, Scotland, 31-35)
\end{quote}
\begin{quote}
    \textit{``Cameron seems more inclined to make mistakes in sensitive topics such as work or personal relationships.''} (P221, F, Spain, 46-50)
\end{quote}
Therefore, deploying probabilistic automation systems in contexts where the impact is assumed lower --- following thorough analysis --- is the only way to increase trustworthiness. In this rationale lies a risk of `judging a book by its cover': rejecting or distrusting systems that provide useful and accurate output simply because it employs probabilistic automation or because it is described in specific terms. 

\paragraph{\textbf{Evaluation factors in the OPSEC rationale}}
The evaluation factors in the OPSEC rationale are somewhat given by the definition of operations security: threats, vulnerabilities, risks, and countermeasures. In lieu of the possibility to thoroughly analyze these factors, the average user will make estimates of \textit{privacy} \cite{baldassarre2024polaris}, \textit{benevolence} \cite{mayer1995integrative} of the system, \textit{risk of failure}, \textit{impact in case of failure} (or ``stakes''), and the \textit{accountability} of the system.


\subsubsection{\textbf{The `wicked world, tame computers' rationale}: Many real world problems are too messy to reduce to algorithmic representation}
A common reasoning in many responses was the conviction that some tasks are more suited for probabilistic automation systems than others, and that many real-world tasks can not be reduced to numbers. The name of this rationale is of course liberally borrowed from Rittel and Webber's exposition of \textit{wicked problems} \cite{rittel1973dilemmas}; planning problems whose nature is messy, and which do not have objectively correct or incorrect solutions. It is the principle of the `wicked world, tame computers' rationale that many problems in the world can not be `represented' accurately enough or with enough detail to be solvable by algorithms. Interestingly, people have different opinions about what \textit{is} appropriate as a problem for probabilistic automation to tackle:
\begin{quote}
    \textit{``I think it would be difficult to determine skin conditions just from a photo.''} (P291, F, Australia, 31-35)
\end{quote}
\begin{quote}
\textit{``Using digital pictures is more accurate than using recorded conversation.''} (P136, F, Japan, 26-30)
\end{quote}
\begin{quote}
    \textit{``There is very little (I think) disagreement about the use of language - words and what they mean''} (P301, F, UK, 56-60)
\end{quote}
\begin{quote}
    \textit{``I feel language is so difficult because of accents and dialects''} (P316, F, UK, 46-50)
\end{quote}
In this rationale, surface numbers are usually preferred to, e.g., text, image, or audio, that is, a probabilistic automation system is often assumed to perform more accurate calculations when the data foundation is numbers on a surface level\footnote{Of course, all probabilistic automation systems compute numbers, so `surface-level' here means, that the data are countable as numbers by humans as well, for instance, money or time.}, than when the data foundation consists of other modalities:
\begin{quote}
    \textit{``Because it uses numbers and statistics that are easier to classify than words.''} (P29, F, Spain, 31-35)
\end{quote}
\begin{quote}
    \textit{``Computers are better with numbers than texts. I would trust more an app with numbers than one who manages texts.''} (P79, M, UK, 36-40)
\end{quote}
Trustworthiness increases if the system is \textit{less complex}, or computes less volatile data, thus decreasing the risk of incorrect or unhelpful output:
\begin{quote}
    \textit{``Dietary tastes are in general reasonably fixed and the app likely has access to more structured, accurate data than the fashion app, where items are constantly changing and decisions are often based on a 'spur of moment' whim''} (P273, M, Italy, 56-60)
\end{quote}
Interestingly, `history' was mentioned by several participants as a factor that supposedly increases the accuracy of the system's output and therefore its perceived trustworthiness, e.g.:
\begin{quote}
    \textit{``I choose the re-Commender because it makes reference to history and then back to the present time.''} (P456, F, UK, 26-30).
\end{quote}
\begin{quote}
    \textit{``I would trust MonAIMaker as it uses history to make predictions.''} (P322, M, South Africa, 31-35)
\end{quote} 
What participants understand by `history' is unclear --- data is always based on the past and is therefore always `history' --- but it illustrates that a person's understanding of the relationship between the past and present (and whether patterns in the past can be used to predict the future) is a determining factor for the person's attribution of trustworthiness to the system. 

Some participants based their evaluation on their judgment of how `developed' or sophisticated the technology or data foundation was:
\begin{quote}
    \textit{``Language learning is more developed in current LLM than science-related areas.''} (P410, M, Spain, 41-45)
\end{quote}
\begin{quote}
    \textit{``I would trust DermAI Scan a little bit more, because I believe that dermatology pictures are more researched than dementia.''} (P167, M, Germany, 31-35)
\end{quote}
Depending on the application context and in contrast to the human favoritism rationale, some respondents also expect that probabilistic automation systems are \textit{better} than humans at many number-crunching tasks, and the inclusion of probabilistic automation technology can therefore improve or augment many processes:
\begin{quote}
    \textit{``I think a jury can do with as much objective help as it can get.''} (P284, M, South Africa, 55-60)
\end{quote}
\begin{quote}
    \textit{``I think AI would be good at collating this, estimating risk etc. It's more mathematical.''} (P70, F, England, 51-55)
\end{quote}
\begin{quote}
    \textit{``I like that MentorMe suggests ways about approaching a problem, rather than solving it.''} (P457, F, Los Angeles, USA, 21-25) 
\end{quote}
\paragraph{\textbf{Consequences of the 'wicked world, tame computers' rationale}}
In the `wicked world, tame computers' rationale, a probabilistic automation system is more likely to be trustworthy if its domain application is perceived to encompass less complex calculations. Trustworthiness of the system also increases if the data foundation is more `specific' or `less complex':
\begin{quote}
    \textit{``I think that the Intellitrade app analyzes news stories - from which news source? I'm a bit apprehensive about trusting news sources, especially with the amount of fake news going around.''} (P352, F, South Africa, 26-30)
\end{quote}
\begin{quote}
    \textit{``I would trust the AquaSentinel AI-MAR more, because I believe it's easier to differentiate objects on the sea than on land.''} (P167, M, Germany, 31-35)
\end{quote}
In practice, the evaluation of what constitutes a more `specific' or `less complex' data foundation appears to be highly subjective --- some respondents believed one type of data was more appropriate to subject to algorithmic calculations and some respondents believed the exact opposite. This discrepancy highlights the knowledge gaps of the general population of what constitutes an accurate representation problem for probabilistic automation technology. It illustrates a risk of people making decisions of trust that are based on faulty assumptions of how probabilistic automation works. People (understandably) have varying knowledge about how algorithms calculate and which data is ``easier'' to analyze and make predictions about. A consequence of this is that trust can potentially be based on the wrong assumptions compared to how an probabilistic automation system actually performs. 

\paragraph{\textbf{Evaluation factors in the 'wicked world, tame computers' rationale}}
The factors that go into the evaluation of trustworthiness in the `wicked world, tame computers' rationale are (assumptions about) the system's \textit{ability} \cite{mayer1995integrative}, the \textit{feasibility} of the promised results, the \textit{type of data} the system uses, and some guess (potentially based on experience or knowledge) about the \textit{sophisticatedness of the algorithms}.


\vspace{2mm}

\begin{table}[]
\small
    \centering
    \renewcommand{\arraystretch}{1.4}%
    \scalebox{0.85}{
    \begin{tabular}{m{2cm}|m{2.4cm}|m{2.6cm}|m{2.6cm}|m{4.6cm}}
         \textbf{Name} & \textbf{Premise} & \textbf{Consequences} & \textbf{Evaluation \newline factors} & \textbf{Examples}  \\ \hline
         
         \textbf{Human \newline favoritism \newline rationale} 
         & A system can not perform as well as a human. 
         & A system is more trustworthy if a(ny) human is in charge of the creation or evaluation of output. 
         & confidence in own and others' ability, degree of human control over the system, passiveness of the system
         & \textit{``I trust human based recommendations better.''} 
         \newline (P128, M, South Africa, 26-30) \vspace{1mm} \newline
         \textit{``there are chances to check the offer taken by the app. So you can check and decide whether to or not.''}
         \newline (P391, M, Germany, 31-35) \vspace{1mm} \newline
         \textit{``to be a true mentor requires very human creativity to tailor response to each individual circumstance.''}
         \newline (P209, F, New Zealand, 61-65)
         \\ \hline

         \textbf{Black box \newline rationale}
         & We should not trust what we can not question and explain.
         & A system is more trustworthy if deployed in contexts where human (domain) experts can evaluate their output.
         & \textit{explainability} \cite{baldassarre2024polaris}, \textit{transparency} \cite{rai2020explainable, von2021transparency}, expertise of end users
         & \textit{``These systems need to be careful that they do not cause undue alarm amongst users.''} 
         \newline (P47, M, Scotland, 56-60) \vspace{1mm} \newline 
         \textit{``Contains more information of how it works''} 
         \newline (P314, F, UK, 51-55) \vspace{1mm} \newline 
         \textit{``This is solely based on the fact that the other operates above sea which I cannot imagine how it would function''} 
         \newline (P3, F, South Africa, 31-35)
         \\ \hline

         \textbf{OPSEC \newline rationale}
         & We should automatically be cautious about any use of `AI'.
         & A system that uses `AI' technology is likely to be prone to some kind of risk. 
         & \textit{privacy} \cite{baldassarre2024polaris}, \textit{benevolence} \cite{mayer1995integrative}, risk of failure, impact in case of failure, accountability, safety, semantics 
         & \textit{``Seems that re-Commender collects less sensitive information.''} 
         (P308, M, Poland, 51-55) \vspace{1mm} \newline
         \textit{``I wouldn't trust AI to invest my money, dining experiences are much lower stakes.''}
         \newline (P298), F, UK, 26-30 \vspace{1mm} \newline
         \textit{``Learning has the potential for bias and I am not going to be educated by a machine.''}
         \newline (P256, F, Spain, 41-45)
         \\ \hline

         \textbf{Wicked world, tame computers rationale}
         & Many real world problems are too messy to reduce to algorithmic representation.
         & A probabilistic automation system is more likely to be trustworthy if its domain application is perceived to encompass less complex calculations. 
         & \textit{ability} \cite{mayer1995integrative}, feasibility, data type, sophisticatedness of algorithms
         & \textit{``MentorMe may not be correct because of the role of substantive knowledge needed''} 
         \newline (P308, M, Poland, 51-55) \vspace{1mm} \newline 
         \textit{``Stock trajectories are not reliable and past stock performance is not an indicator for future.''}
         \newline (P293, M, Italy, 26-30) \vspace{1mm} \newline
         \textit{``Driverless is still a distance off. The trolley problem can be an issue.''}
         \newline (P114, M, South Africa, 31-35)
         \\ \hline

    \end{tabular}}
    \caption{An overview of the four rationales identified in the data. The \textbf{Premise} is the foundation or the logical assumption that the rationale is based on. The \textbf{Consequences} are the impact to how trustworthiness is ascribed to a system. The \textbf{Evaluation factors} are features of the system that influence the evaluation of the system as more or less trustworthy. The terms marked in \textit{italics} are terms from previous literature on technology acceptance and trustworthiness -- they are listed along with their reference.}
    \label{tab:rationales}
\end{table}

\noindent A summary and overview of the rationales is presented in Table \ref{tab:rationales}. The point of these rationales is to provide a distilled version of a logical reasoning which may be more or less present in someone's evaluation of trustworthiness of a probabilistic automation system. One person could experience all rationales in their evaluation and decision making, but one rationale might weigh higher than others in certain situations or for certain systems.


\section{Discussion, limitations, and future work}

Explainable `AI' (`XAI') entails \textit{``codifying and reporting on plans and actions in a way that is intelligible, relevant, and intriguing to humans''} \cite{carroll2022should}. One of the key elements of developing `XAI' is assessing  users' trust and reliance in the system, and whether these are appropriate \cite{hoffman2018metrics}. Mapping and understanding (latent) motivations and rationales makes sense in a domain where extensive public discourse is bound to shape people's opinion a priori (meaning, here, before people actually experience these systems first hand). A general black boxing of these systems combined with lack of inclusion of users in the development of the systems means that many people experience (at the same time) both excitement about the expansion of their own creative potential, and fear of losing their jobs \cite{inie2023designing}. The rationales presented in this paper offer a language for us to discuss the lines of reasoning that motivate people's trust in probabilistic automation, and an empirical account of the presence of different rationales in the participant population of this survey. 

\subsection{Comparison of the rationales with what we already knew}
While many of the reasons and rationales provided by the participants' statements correspond almost directly with factors and criteria identified by literature on technology acceptance and `AI' trustworthiness (for instance, factors such as \textit{explainability}, \textit{privacy} \cite{baldassarre2024polaris}, \textit{perceived usefulness} \cite{davis1989perceived}, \textit{ability}, and \textit{benevolence} \cite{mayer1995integrative}), some patterns from this empirical study are distinctively different from theoretical or literature survey-based findings -- or theoretical criteria can be elaborated with respect to the specific case of probabilistic automation systems.

\subsubsection{The human favoritism bias extends to the fundamental judgment of a probabilistic automation system}
First and foremost, this study confirms and expands upon the \textit{human favoritism} bias coined by Zhang and Gosline \cite{zhang2023human}. In their study, the bias applied to the evaluation of quality of output from a generative `AI' model. From the results of the current study, the bias appears to extend beyond the evaluation of the quality of output to the fundamental judgment or evaluation of a probabilistic automation system as a whole -- none of the participants in the current study had ever seen any output of any of the systems they were evaluating, but many still described the human factor and human control as essential for their willingness to trust the system. Considering the differences between the two rationales \textit{human favoritism} and \textit{black box}, it would be interesting to examine further the differences between the influence of human control versus human knowledge or expertise. Potentially, human trust in `experts' could mean that people trust experts' use of probabilistic automation systems without knowing if the systems exacerbate or reinforce human errors, biases, and mistakes. The human favoritism bias is important to investigate further, and could provide valuable insights for the (re)design of human-in-the-loop-protocols \cite{zhang2023human}.

\subsubsection{Fairness and ethics are almost non-existent terms in the dataset}
Despite extensive scholarly and popular media focus on concepts such as \textit{fairness} \cite{baldassarre2024polaris, mehrotra2024integrity}, \textit{integrity}\cite{mayer1995integrative, mehrotra2024integrity}, and \textit{ethics} \cite{hagendorff2020ethics, birhane2021algorithmic, bender2021dangers} in relation to probabilistic automation systems, these terms are largely non-existent in our dataset. A quick search on Google Scholar yields more than 25,000 hits for the search ``AI fairness'' and more than 21,000 hits for the search ``AI ethics'' only since 2023 (approximately one year), but the terms ``fair'' and ``ethic[]'' are only present \textit{three} times each among the more than 3000 survey responses (and primarily in relation to the evaluation of the legal recommendations systems). Copyright issues and and lack of transparency of training data are also heavily discussed in scholarly and public debate, but the term ``copyright'' is mentioned zero times in the data from this study (plagiarism is mentioned once in relation to the consequence for the user of using the MentorMe system to help with their academic work). 

This does not necessarily mean that the concepts are not present in people's minds, but it does suggest that either the terms are not the most relevant ones to people's evaluation of trust and trustworthiness, \textit{or} that the concepts are relevant, but the specific terms are not part of popular vernacular. The extensive focus on \textit{risk}, \textit{privacy}, and \textit{impact in case of failure} under the OPSEC rationale certainly illustrates deliberations about ethics of these systems, even if the specific terms are not present. 

\subsubsection{Security or safety?}
Baldassare et al. \cite{baldassarre2024polaris} defined \textit{security} --- one of the four core principles of `Trustworthy AI' --- broadly as \textit{technical robustness}, encompassing \textit{safety}, and \textit{non-maleficence}, mapping to system requirements like `resilience to attack, `fallback plans', `general safety', `accuracy', `reliability', and `reproducibility'. For systems development, this is a fairly infertile definition of security due to the level of abstraction and amount of concepts crowded under one umbrella. \textit{Risk} in general is not easily assessed in the deployment of probabilistic automation systems, and especially not with the use of natural language based interaction \cite{derczynski2023assessing}. It appears from the results of this data, though, that risk is highly present in people's conscious reasoning about trustworthy probabilistic automation. One pertinent question is how to assess risks in terms of security (of the system) versus safety (of people). A common theme in the OPSEC rationale was choosing a system that did not impose risk of damage to people over one that would. On a surface level, this appears as obvious reasoning, but a principle that will be challenged in practice when technology is deployed. First, a choice is easier if the damage is \textit{overt}, and more difficult if the risk of damage is not directly foreseeable. The participants mentioning damage or safety were most often doing so under the evaluation of the autonomous vehicles or military drones, and less often (at least on a surface level) under the other system pairs. Second, the autonomous vehicles and military drones also carry a risk to \textit{physical} safety, which appears to take higher priority than, for instance, unfair or biased algorithms in favor of certain recommendations than others (in case of the recommender systems). Safety in terms of other factors were not as present in evaluations --- for instance the safety of the environment. Third, the it appears that the risk most present in people's rationales is `impact in case of failure' (what could happen if the system fails) rather than actual `risk of failure' (how likely is the system to fail). At the extreme of this reasoning, if a system fails 20\% of the time but the impact is low, would it be considered more trustworthy than a system which fails 0.01\% of the time but carries a risk of physical damage to people? Which people are more prone to be subject to the impact of failure, and are there types of damage that is more acceptable than others? In light of such questions, we might ask ourselves: is it then the only \textit{security} that should be a principle of trustworthy `AI', or should such principles include \textit{safety} --- ensuring that people, animals, the environment, are safe --- when the system is deployed?

\subsubsection{Claims are strong, support varies substantially, and warrants are abstruse}
According to a standard simplified  model of reasoning, a person who makes a claim usually uses some support (grounds), which is based on a warrant that links the grounds to the claim \cite{karbach1987using}. Acknowledging that it is difficult to construct a full, detailed model of a person's support and warrant (let alone potential qualifiers, backing, or rebuttals) for their reasoning based on a survey response, it is clear from the collective data that many strong claims exist about the trustworthiness of probabilistic automation systems, the people who deploy it, and the people use it. The support --- the grounds on which these claims and opinions are based --- varies substantially between people. From this study alone, numerous misunderstandings or gaps in knowledge could be identified relatively easily, and in some responses, participants appeared to fill in some arbitrary blanks of the systems descriptions that could lead to worrying misunderstanding in a real-life scenario (such as the respondents who assumed a bus driver would be present in the Commuter vehicle, when no such driver was mentioned in the description). Furthermore, it is not clear on which warrants people base their reasoning. Many respondents would make directly opposing arguments for why they assumed one system would be more reliable or accurate than the other. Considering the black box nature of probabilistic automation systems, it is worth thinking about appropriate ways of educating the general public to understand these technologies in ways that can help them formulate support and warrants for their assumptions --- or at least in a way that allows people to communicate from common grounds about the systems --- to avoid more polarization about the topic than absolutely necessary.

\vspace{2mm}

\noindent The contributions to known research described above suggest that we might use empirical explorative research such as the data presented in this paper to uncover ``blind spots'' in our research about trust in and trustworthiness of probabilistic automation systems. Important factors for many users may not be important for all, and factors may be important for different reasons. Empirical research such as this also help illustrate dissonances between the picture painted by scholarly reports or popular media, and the picture present in the broad population. Some limitations of this work and suggestions for how it could be expanded upon are discussed below.

\subsection{Limitations and future work}
The first limitation of this work is imposed by the mode of study: survey responses are bound to provide fewer details about emotions and cognition than, for instance, qualitative interviews. Of course, many tacit rationales could be present in someone's reasoning without being expressed in something as simplified as a survey response. As described in the methodology section, this was decided a trade-off between reaching a broad range of people from many different countries and depth of data, and it means that different modes of investigation might elicit deeper and, perhaps, different rationales than the ones uncovered in this study. As in any line of research, since these rationales are generated based on inductive analysis, it would make sense to explore and test (the limits of) the rationales in deductive practice with people who can explain their thinking in greater depth. 

A second limitation is that the study is based on an arbitrary -- but not random -- sample of survey participants from the platform Prolific, and the results are not guaranteed to generalize to all populations. There is an over-representation of participants from Europe and South Africa which is due to the sampling from Prolific. Therefore, the rationales probably do not uncover all or even necessarily the most prevalent motivations or rationales that exist for trusting probabilistic automation systems, only the most apparent ones in this data, and only the ones that participants chose to express. This is also the reason this research does not quantify the presence of different codes or rationales in different groups of people. The goal of this study was not to draw general conclusions about everyone's rationales, neither to quantify rationales in terms of importance, but rather to uncover some of the existing dynamics at play. It would be interesting to explore the presence or dominance of different rationales or motivations in different groups of people. It is also likely that certain rationales are more present when evaluating different groups of products or systems -- for instance, it is likely that products that are deployed with direct responsibility of human lives, such as autonomous vehicles or military drones, elicit more cautionary rationales and more conservative evaluations than products which have ``lower stakes'' (in the words of several survey participants).

\subsection{Ethical statement}
The online survey design adhered to the ethical guidelines in HCI methodology \cite{bruckman2014research} to guarantee participant anonymity and data privacy. Participants were recruited via the Prolific platform, and compensated for their participation. To ensure the participants constituted a representative group, they were screened by criteria such as country of residence, self-assessed socio-economic status, ethnicity, and geographic location. No identifiable information was collected, and all the survey responses were stored temporarily on a secure server. To avoid confusion about the fictitious products, a statement was added at the end of the survey asserting that all products are 100\% imagined, although some of them have been loosely based on existing products or services. The debriefing text also stated that the goal of the research was to investigate whether the description of the product influenced the way its trustworthiness and functionality is perceived, as well as author contact info.

While the study addresses a timely question of why people trust automation driven systems, it could also lead to a potential dual use. Bad actors could use the results to elicit unearned trust from people, for instance by manipulating their systems or descriptions of their systems to appear to appear to adhere to the rationales more than they did in reality. This analysis is not presented with that intention, but only for the purpose of increasing our knowledge about the interplay between motivation, rationales, and trust by people from the broad public. The results are intended to be used to inform developers and designers about the potential risks of trust in, reliance on, and skepticism of probabilistic automation systems.


\section{Conclusions}
This paper presented the results of a qualitative analysis of a broad group of people's responses to questions of trust in probabilistic automation (or `AI') systems. More than 450 participants were presented with eight pairs of probabilistic automation systems and asked to choose which one they would trust more, either as a user themselves, or to provide better output in general. The more than 3000 written responses were analyzed using thematic analysis. The themes derived were \textit{incentives} and \textit{counterincentives} for using probabilistic automation systems, as well as four rationales (reasons for ascribing more or less trustworthiness to a system): the \textit{Human favoritism rationale}, the \textit{Black box rationale}, the \textit{OPSEC rationale}, and the \textit{`Wicked world, tame computers' rationale}. These rationales distill lines of reasoning for assessing the trustworthiness of a probabilistic automation system, and provide empirical contributions to current research about probabilistic automation and trustworthiness.

\section*{Supporting information}

\paragraph*{S1 Fig.}
\label{codingexample}
{\bf Coding example from the initial thematic analysis.}

\paragraph*{S2 Fig.}
\label{age}
{\bf Participants by age.}

\paragraph*{S3 Fig.}
\label{gender}
{\bf Participants by gender.}

\paragraph*{S4 Fig.}
\label{continent}
{\bf Participants by continent.}

\section*{Acknowledgments}
Thank you so much to Stefania Druga, Emily M. Bender, and Peter Zukerman for the co-design of this study and the continuous collaborations.


%
%
%
(\bibliography{references})

\clearpage

\section{Product descriptions}

\begin{table}[h!]
\small
    \centering
    \renewcommand{\arraystretch}{1.4}%
    \scalebox{0.8}{
    \begin{tabular}{m{0.8cm}|m{3.6cm}|m{3.6cm}|m{3.6cm}|m{3.6cm}}
         \textbf{Name} & \textbf{Description 1} & \textbf{Description 2} & \textbf{Description 3} & \textbf{Description 4}  \\ \hline
         
         \rotatebox{90}{\textbf{reCommender}} 
         & A smartphone app, re-Commender, understands your dining preferences, knows your preferences from historical data, and uses trends from all its users to predict new restaurants you might enjoy. It remembers your previous ratings and habits, and figures out offers and coupons you might like.
         & A smartphone app, re-Commender, creates a model of your dining inclinations, encodes your preferences from historical data, and uses trends from all its users to classify new restaurants you might enjoy. It stores your previous ratings and habits, and extracts offers and coupons you might like. 
         & A smartphone app, re-Commender, collects data about your dining experiences. It analyzes your preferences from historical data, and uses trends from all of its users to choose new restaurants you might enjoy. It stores your previous ratings and habits, and picks offers and coupons you might like.
         & A smartphone app, re-Commender, is programmed to collect data about your dining experiences. You can use it to analyze your preferences from historical data, and trends from all of its users to get suggestions for new restaurants you might enjoy. You can save your previous ratings and habits, and find offers and coupons you might like.
         \\ \hline

        \rotatebox{90}{\textbf{IntelliTrade}}
        & A smartphone app, IntelliTrade is an intelligent stock broker, which identifies promising stocks, funds, and bonds for you. It remembers your investment preferences as well as historical stock trajectories and understands current news stories, using both of these to predict promising investment opportunities.
        & A smartphone app, IntelliTrade is an automated stock broker, which makes calculations about promising stocks, funds, and bonds for you. It encodes your investment preferences as well as historical stock trajectories and processes current news stories, using both of these to classify promising investment opportunities.
        & A smartphone app, IntelliTrade is a personal stock broker, which identifies promising stocks, funds, and bonds. It collects data about your investment preferences as well as historical stock trajectories. It also analyzes current news stories, using these to select promising investment opportunities.
        & A smartphone app, IntelliTrade is a personal stock broker, which you can use to identify promising stocks, funds, and bonds. It is programmed to store data about your investment preferences as well as historical stock trajectories. The algorithms are also frequently run over current news stories, so you can use them to find promising investment opportunities.
        \\ \hline

        \rotatebox{90}{\textbf{MonAIMaker}}
        & MonAIMaker is an intelligent app that helps you plan your personal finances. It learns what you are likely to spend money on by recognizing trends in your bank statements as well as your email correspondences. It uses these to identify ways to save money, and remember when you have bills and expenses due.
        & MonAIMaker is an automatic pattern matching app that helps you plan your personal finances. It classifies what you are likely to spend money on by mapping trends in your bank statements as well as your email correspondences. It uses these to provide suggestions for ways to save money, and store information about when you have bills and expenses due.
        & MonAIMaker is an app that helps you plan your personal finances. It monitors what you are likely to spend money on by identifying trends in your bank statements as well as your email correspondences. It uses these to find ways to save money and remind you when you have bills and expenses due.
        & MonAIMaker is an app that you can use to plan your personal finances. It is programmed to monitor what you are likely to spend money on based on calculations of trends in your bank statements as well as your email correspondences. The data can be used to find ways to save money and to set up reminders when you have bills and expenses due.
        \\ \hline

        \rotatebox{90}{\textbf{Cameron}}
        & An app, Cameron, is powered by artificial intelligence and machine learning to help you organize and answer your emails. It interprets text from your incoming emails, suggests answers based on your writing style, and recognizes tasks and deadlines to create automated to-do-lists for you.
        & An app, Cameron, is powered by automatic pattern matching and machine conditioning to help you organize and answer your emails. It classifies text from your incoming emails, synthesizes answers based on your writing style, and assigns labels to tasks and deadlines to create automated to-do-lists for you.
        & An automatic app, Cameron, helps you organize and answer your emails. It classifies text from your incoming emails, and it suggests answers based on your writing style. It also identifies tasks and deadlines to create automated to-do-lists for you.
        & An automatized  app, Cameron, is a system you can use to organize and answer your emails. It is programmed to classify text from your incoming emails, and you can use it to generate answers based on your writing style. You can also use it to identify tasks and deadlines to create automated to-do-lists.
        \\ \hline

    \end{tabular}}
    \vspace{1mm}
    \caption{Product descriptions 1-4 as they were shown to participants. To mitigate language effects, four different descriptions of each system was created and all participants were randomly shown one of the four.}
    \label{tab:productdescriptions1}
\end{table}

\begin{table}[]
\small
    \centering
    \renewcommand{\arraystretch}{1.4}%
    \scalebox{0.8}{
    \begin{tabular}{m{0.8cm}|m{3.6cm}|m{3.6cm}|m{3.6cm}|m{4cm}}
         \textbf{Name} & \textbf{Description 1} & \textbf{Description 2} & \textbf{Description 3} & \textbf{Description 4}  \\ \hline

        \rotatebox{90}{\textbf{HaulIT}}
        & A self-driving truck HaulIT handles long haul freight 24/7 without rest stops, and it never gets tired or distracted. The truck is designed for both city and highway, meaning it always chooses the most optimal route for speed and efficiency by analyzing current and projected traffic conditions and self-managing battery charging.
        & A driverless truck HaulIT is programmed to transport long haul freight 24/7 without rest stops, and it never gets tired or distracted. The truck is designed for both city and highway, meaning it is always sent along the most optimal route for speed and efficiency, based on statistical predictions about current and projected traffic conditions as well as optimal battery charging points.
        & A neural networks-based truck HaulIT is made for long haul freight 24/7 without rest stops. The constantly evolving algorithms use a metabolic principle to minimize their own synaptic activity (cost) while maximizing their impact, meaning the truck uses the most efficient routes in both cities and on highways, based on neural predictions about traffic conditions and optimal battery charging points.
        & A weighted networks-based truck HaulIT is made for long haul freight 24/7 without rest stops. The constantly updated algorithms use an optimizing principle to minimize their own computing activity (cost) while maximizing their impact, meaning the truck uses the most efficient routes in both cities and on highways, based on weighted node-predictions about traffic conditions and optimal battery charging points.
        \\ \hline

        \rotatebox{90}{\textbf{Commuter}}
        & A sleeper bus, Commuter, drives people from their home to a long-distance destination overnight. The bus avoids other vehicles and obstacles on the road, and adapts to the weather conditions to navigate safely. It monitors traffic live and picks the best and safest routes.
        & A sleeper bus, Commuter, is used to transport people from their home to a long-distance destination overnight. The bus has algorithms for avoiding other vehicles and obstacles on the road, and the algorithms are adjusted to the weather conditions to navigate safely. Its systems are fed live traffic data for calculations of the best and safest routes.
        & A driverless sleeper bus, Commuter, is programmed with neural networks to transport people from their home to a long-distance destination overnight. The bus is equipped with swarm intelligence to avoid other vehicles and on the road, and the neural network adjusts to weather conditions to navigate safely. Its artificial synapses are constantly fed live traffic data for calculations of the best and safest routes.
        & A driverless sleeper bus, Commuter, is programmed with weighted networks to transport people from their home to a long-distance destination overnight. The bus is equipped with optimization algorithms to avoid other vehicles and on the road, and the weighted network adjusts to weather conditions to navigate safely. Its network nodes are constantly input live traffic data for calculations of the best and safest routes.
        \\ \hline

        \rotatebox{90}{\textbf{AquaSentinel AI-MAR}}
        & An AI and ML-powered drone, AquaSentinel AI-MAR, monitors enemy seas. Armed with cutting-edge technology, it autonomously patrols waterways, utilizing advanced algorithms to swiftly detect and analyze potential threats in real-time.
        & An AI and ML-powered drone, AquaSentinel AI-MAR, is programmed to monitor enemy seas. Armed with cutting-edge technology, it is positioned over  waterways, equipped with advanced algorithms designed to detect and provide analyses of potential threats in real-time.
        & An neural network-powered drone, AquaSentinel AI-MAR, is programmed to passively monitor enemy seas. Equipped with advanced digital  senses, it is watching and listening over waterways. Its neural network has been designed to detect and provide analyses of potential threats in real-time.
        & An weighted network-powered drone, AquaSentinel AI-MAR, is programmed to passively monitor enemy seas. Equipped with advanced digital  sensors, it is recording video and sound over waterways. Its weighted network has been designed to detect and provide analyses of potential threats in real-time.
        \\ \hline

        \rotatebox{90}{\textbf{AI Scan Guards}}
        & The newest unmanned aircraft systems (UAS), AI Scan Guards, monitor a physical territory from the air. They use image recognition to analyze live video streams, seek out enemy targets and alert the defense forces.
        & The newest unmanned aircraft systems (UAS), AI Scan Guards, are programmed to monitor a physical territory from the air. They are equipped with image recognition algorithms that are used to process live video streams. System outputs may be used to identify enemy targets and provide alerts to defense forces.
        & The newest unmanned aircraft systems (UAS), AI Scan Guards, use neural networks to passively watch a physical territory from the air. Their neural networks are specifically trained on image recognition tasks with live video streams, meaning they see and hear activity instantly.
        & The newest unmanned aircraft systems (UAS), AI Scan Guards, use weighted networks to passively record video of a physical territory from the air. Their weighted networks are specifically trained on image recognition tasks with live video streams, meaning the predictions identify image and sound activity instantly.
        \\ \hline

    \end{tabular}}
    \caption{Product descriptions 5-8 as they were shown to participants. To mitigate language effects, four different descriptions of each system was created and all participants were randomly shown one of the four.}
    \label{tab:productdescriptions2}
\end{table}

\begin{table}[]
\small
    \centering
    \renewcommand{\arraystretch}{1.4}%
    \scalebox{0.75}{
    \begin{tabular}{m{0.8cm}|m{3.6cm}|m{3.6cm}|m{3.6cm}|m{4cm}}
         \textbf{Name} & \textbf{Description 1} & \textbf{Description 2} & \textbf{Description 3} & \textbf{Description 4}  \\ \hline

        \rotatebox{90}{\textbf{Judy}}
        & A software system for court juries, Judy, uses neural networks to inform jury members in court cases. Thousands of transcripts and outcomes from previous similar trials are fed to Judy’s brain, whose digital neurons digest all data and determinants to provide information about relevant law and precedence in current 
        & A software system for court juries, Judy, uses weighted networks to inform jury members in court cases. Thousands of transcripts and outcomes from previous similar trials are input into Judy’s CPU, whose algorithms process all data and determinants to provide information about relevant law and precedence in current 
        & A conversational software system for court juries, Judy, can be used by jury members to discuss active court cases. Judy is based on thousands of transcripts and outcomes from previous similar trials and can tell the jury about complex law and precedence. The jury can ask Judy to process any kind of data and to suggest further avenues of research. All use of Judy is disclosed openly in  court.
        & A generative software system for court juries, Judy, can be used by jury members to gather information about active court cases. Judy is based on thousands of transcripts and outcomes from previous similar trials and can produce text for the jury about complex law and precedence. The jury can input any kind of data into Judy to process and produce output candidate matches for further avenues of research. All use of Judy is disclosed openly in  court.
        \\ \hline

        \rotatebox{90}{\textbf{JurisDecide}}
        & A software application, JurisDecide uses neural networks to enhance lawyers’ decision-making in trials. Its digital brain continually evolves and rapidly processes extensive legal data, including precedent and case law which JurisDecide’s digests to spit out information for legal professionals. 
        & A software application, JurisDecide uses weighted networks to enhance lawyers’ decision-making in trials. Its algorithms continually self-update and rapidly process extensive legal data, including precedent and case law which JurisDecide’s processes to output information for legal professionals. 
        & A generative AI application, JurisDecide  can be used by lawyers during trials to speak to and debate their own decision-making processes. JurisDecide is both a source of information and a chatbot: it rapidly processes extensive legal data, including precedent and case law, and can both ask questions of and answer questions from legal professionals.
        & A generative AI application, JurisDecide can be used by lawyers during trials to input speech and think out loud about their own decision-making processes. JurisDecide is both a source of information and a generative text software: it rapidly processes extensive legal data, including precedent and case law, and can both produce text in the form of questions and answers for legal professionals.
        \\ \hline

        \rotatebox{90}{\textbf{MindHealth}}
        & A neural network system, MindHealth, is an online digital ear which senses indicators in spoken language that a person may be developing one or more early signs of dementia. Its digital synapses have evolved during thousands of conversations with healthy humans and dementia patients.
        & A weighted network system, MindHealth, is an online digital recorder which classifies indicators in spoken language that a person may be developing one or more early signs of dementia. Its complex algorithms have been fine-tuned based on thousands of conversations with healthy humans and dementia patients.
        & MindHealth is an online digital conversation partner, which you can talk to via your own computer. It classifies indicators in spoken language and can tell you if you may be developing one or more early signs of dementia. Its complex algorithms have been fine-tuned based on thousands of conversations with healthy humans and dementia patients, and you can ask it questions about its assessment and have it suggest further routes of investigation.
        & MindHealth is an online digital recorder, which you can input speech to via your own computer. It classifies indicators in spoken language and can output a statistical prediction about whether you may be developing one or more early signs of dementia. Its complex algorithms have been fine-tuned based on thousands of conversations with healthy humans and dementia patients, and you can input questions about its assessment and have it output text about further routes of investigation.
        \\ \hline

        \rotatebox{90}{\textbf{DermAI Scan}}
        & A diagnostic tool, DermAI Scan, uses neural networks to diagnose dermatological conditions from your home computer. You feed it a picture and receive a suggestion for a diagnosis. Evolving neural networks means that the system’s neurons can instantly compare your picture to images of millions of previous diagnoses.
        & A diagnostic tool, DermAI Scan, uses weighted networks to diagnose dermatological conditions from your home computer. You upload a picture and receive a suggestion for a diagnosis. Fine-tuned weighted networks means that the system’s weights can instantly compare your picture to images of millions of previous diagnoses.
        & DermAI Scan uses AI to respond to an uploaded picture with suggestions for potential dermatological conditions. Fine-tuned algorithms instantly compare your picture to images of millions of previous diagnoses and tell you the likelihood of different ones. It can discuss different possible diagnoses with you if you tell it more about the history of your condition. 
        & DermAI Scan uses AI to generate statistical predictions about potential dermatological conditions based on an uploaded picture. Fine-tuned algorithms instantly compare your picture to images of millions of previous diagnoses and output the likelihood of different ones. It can generate text about different possible diagnoses if you input more about the history of your condition. 
        \\ \hline

    \end{tabular}}
    \caption{Product descriptions 9-12 as they were shown to participants. To mitigate language effects, four different descriptions of each system was created and all participants were randomly shown one of the four.}
    \label{tab:productdescriptions3}
\end{table}

\begin{table}[]
\small
    \centering
    \renewcommand{\arraystretch}{1.4}%
    \scalebox{0.8}{
    \begin{tabular}{m{0.8cm}|m{3.6cm}|m{3.6cm}|m{3.6cm}|m{3.6cm}}
         \textbf{Name} & \textbf{Description 1} & \textbf{Description 2} & \textbf{Description 3} & \textbf{Description 4}  \\ \hline

        \rotatebox{90}{\textbf{Lingua}}
        & A smartphone app, Lingua, is an interactive language learning tutor. You can talk or write to the app and it will speak back to you in real time. Lingua tells you about the accuracy and complexity of your speech, and it suggests areas of improvement.
        & A smartphone app, Lingua, is an interactive language learning tutor. You can input speech or text to the app and it will output speech to you in real time. Lingua indicates the accuracy and complexity of your speech, and it produces suggestions for areas of improvement.
        & A machine learning-based app, Lingua, is an intelligent language learning tutor. It understands both speech and text and it will produce answers to you in real time. Lingua identifies the accuracy and complexity of your speech, and it recognizes areas of potential improvement in your spoken language.
        & A machine conditioning-based app, Lingua, is an automated language learning tutor. It processes both speech and text and it will produce answers to you in real time. Lingua encodes the accuracy and complexity of your speech, and it classifies areas of potential improvement in your spoken language.
        \\ \hline

        \rotatebox{90}{\textbf{MentorMe}}
        & MentorMe is an online chatbot, which you can talk to about specific academic topics (each based on different data sets). It speaks like a mentor, and proposes new ways to approach a problem, rather than just answering questions directly. It also asks you questions to enhance your learning about a given topic.
        & MentorMe is an online chatbot, into which you can input text about specific academic topics (each based on different data sets). It produces text in the style of a mentor, and outputs candidate matches for new ways to approach a problem, rather than just indicating answers for questions directly. It also outputs questions to enhance your learning about a given topic.
        & MentorMe is an intelligent online chatbot, with extensive knowledge about specific academic topics (each based on different data sets). It understands topic-specific questions, and imagines new ways to approach a problem, rather than just answering questions directly. It also comes up with questions to enhance your learning about a given topic.
        & MentorMe is an automated online chatbot, with extensive data about specific academic topics (each based on different data sets). It processes topic-specific questions, and generates text suggesting new ways to approach a problem, rather than just answering questions directly. It also produces questions to enhance your learning about a given topic.
        \\ \hline

        \rotatebox{90}{\textbf{WardrobeEase}}
        & A smartphone app, WardrobEase, is a service for effortlessly restocking essential clothing items such as jeans, socks, and underwear. It discusses your fabric and style preferences with you, you tell it your sizes, and it responds with pictures of choices. You can tell it when your clothes are starting to wear out, and ask it to recurringly order new items from your favorite stores ahead of time.
        & A smartphone app, WardrobEase, is a service for effortlessly restocking essential clothing items such as jeans, socks, and underwear. It allows you to record and specify your fabric and style preferences, you input your sizes, and it outputs pictures of choices. You can mark when your clothes are starting to wear out, and input automatic, recurring orders of new items from your favorite stores ahead of time.
        & A phone app, WardrobEase, is an artificial intelligence-based service for effortlessly restocking essential clothing items such as jeans, socks, and underwear. It will learn your sizes and fabric preferences, and suggest pictures of style choices. It predicts when clothes are likely to wear out, and can be instructed to remember to order new items from your favorite stores ahead of time.
        & A phone app, WardrobEase, is an automatic pattern matching service for effortlessly restocking essential clothing items such as jeans, socks, and underwear. It will encode your sizes and fabric preferences, and display pictures of style choices. It makes statistical calculations about when clothes are likely to wear out, and can be instructed to automatically order new items from your favorite stores ahead of time.
        \\ \hline

        \rotatebox{90}{\textbf{Shoppr}}
        & A smartphone app, Shoppr, lets you create meal plans by discussing your dietary wishes with you. You can tell the system about constraints of health, time, nutrition, budget, and it responds with suggestions for meals, as well as write a meal plan with recipes and order groceries online for you.
        & A smartphone app, Shoppr, lets you create meal plans based on your dietary wishes. You can input constraints of health, time, nutrition, budget into the system, and it produces suggestions for meals, as well as generate a meal plan with recipes and an option to put in an online order for groceries.
        & A smart app, Shoppr, lets you create meal plans based on your dietary wishes. It can remember constraints about health, time, nutrition, and budget, and identify ideas for meals. It can imagine monthly meal plans with recipes and recognize when to order groceries online.
        & A phone app, Shoppr, lets you create meal plans based on your dietary wishes. It can encode constraints about health, time, nutrition, and budget, and synthesize ideas for meals. It can produce monthly meal plans with recipes and make statistical predictions about when to order groceries online.
        \\ \hline

    \end{tabular}}
    \caption{Product descriptions 13-16 as they were shown to participants. To mitigate language effects, four different descriptions of each system was created and all participants were randomly shown one of the four.}
    \label{tab:productdescriptions4}
\end{table}

\end{document}